\documentclass{aa}  
\usepackage{graphics}
\usepackage{txfonts}
\usepackage{longtable}

\begin{document}

\title{Hen~2--104: A close-up look at the Southern Crab\thanks{Based on
observations with the NASA/ESA Hubble Space Telescope, obtained at the 
Space Telescope Science Institute, which is operated by the Association of 
Universities for Research in Astronomy, Inc.~ under NASA contract No.~NAS5-26555; 
on observations obtained at the 4~m~NTT and the 8~m~VLT telescopes of
the European Southern Observatory in Chile.}}


\author{M. Santander-Garc\'\i a\inst{2,1}
	\and R. L. M. Corradi\inst{2,1}
	\and A. Mampaso\inst{1}
	\and C. Morisset\inst{3}
	 \and U. Munari\inst{4}
	\and M. Schirmer\inst{2}
	 \and B. Balick\inst{5}
	 \and M. Livio\inst{6}
	}

\institute{
         Instituto de Astrof\'\i sica de Canarias, 38200 La Laguna,
        Tenerife, Spain
         \\e-mail: miguelsg@iac.es; amr@iac.es
        \and
         Isaac Newton Group of Telescopes, Ap.\ de Correos 321,
         38700 Sta. Cruz de la Palma, Spain
         \\e-mail: msantander@ing.iac.es; rcorradi@ing.iac.es; mischa@ing.iac.es
        \and
        Instituto de Astronom\'\i a, Universidad Nacional Aut\'onoma de M\'exico, M\'exico, D.\ F., M\'exico
        \\e-mail: morisset@astroscu.unam.mx
        \and
        INAF Osservatorio Astronomico di Padova, via dell'Osservatorio 8,
        36012 Asiago (VI), Italy
        \\e-mail:munari@pd.astro.it
         \and
         Department of Astronomy, University of Washington, Seattle, Washington 98195-1580, USA
         \\e-mail: balick@astro.washington.edu
        \and
        Space Telescope Science Institute (STScI), 3700 San Martin Drive,
         Baltimore, MD 21218, USA 
        \\e-mail: mlivio@stsci.edu
        }

\offprints{M. Santander-Garc\'\i a}

\date{\today}

\abstract
{}
{The kinematics, shaping, density distribution, expansion distance,
and ionized mass of the nebula Hen~2--104, and the nature of its
symbiotic Mira are investigated.}
{A combination of multi-epoch HST images and VLT integral field
high-resolution spectroscopy is used to study the nebular dynamics
both along the line of sight and in the plane of the sky. These
observations allow us to construct a 3-D spatio-kinematical model of
the nebula, which together with the measurement of its apparent
expansion in the plane of the sky over a period of 4 years, provides
the expansion parallax for the nebula. The integral field data
featuring the [S{\sc ii}] \ $\lambda\lambda$671.7,673.1 emission line
doublet provide us with a density map of the inner lobes of the
nebula, which together with the distance estimation allow us to
estimate its ionized mass.}  
{We find densities ranging from n$_e$=500 to 1000 cm$^{-3}$ in the
inner lobes and from 300 to 500cm$^{-3}$ in the outer lobes. We
determine an expansion parallax distance of 3.3$\pm$0.9~kpc to
Hen~2--104, implying an unexpectedly large ionized mass for the nebula of the
order of one tenth of a solar mass.}
{}


\authorrunning{Santander-Garc\'\i a et al.}
\titlerunning{A close up to Hen~2--104}
\maketitle

\section{Introduction}

Symbiotic stars are valuable laboratories for studying the physics
of a variety of astrophysical processes, like
mass transfer, accretion and outflow, thermonuclear runaways,
photo-ionization and shock-ionization, and the physico-chemistry
of gas and dust in complex environments containing both hot and cold
stars.

The formation of the nebulae that surrounds some of them, especially
those containing a Mira variable, is one example of these complex
interactions, and the key to understand crucial aspects of the mass
loss history and geometry from these systems. A small fraction of the
Mira wind is believed to be accreted by the white dwarf companion
(Kenyon \& Webbink \cite{KW84}), resulting in quasi-stable
thermonuclear burning at its surface (e.g. Sokoloski \cite{So02})
which sometimes ends in thermonuclear outbursts (e.g. Kenyon
\cite{Ke86}, Livio et al. \cite{Li89}) lasting for hundreds of years
(e.g. Munari \cite{Mu97}). The fast winds from the white dwarf collide
with the slower, non-accreted Mira wind, and shape complex
circum-binary nebulae.
Currently, about a dozen of these nebulae have been
 optically resolved (Corradi \cite{Co03}).
Most of them show bipolar morphologies,
similar to those of many planetary nebulae (PNe) and protoplanetary nebulae
(PPNe) (Corradi \& Schwarz \cite{Co95}, Sahai \cite{Sa02}).

The distance of these system is generally poorly known, in spite of
being a basic parameter to determine their overall physical
properties, like their total luminosities, the mass of their ejecta or
their kinematical ages. Recently, the advent of high spatial
resolution instruments like the HST allows to face this problem by
measuring the expansion parallax of some of these nebulae. The
procedure is to measure the growth of the nebula in the plane of the
sky via multi-epoch imaging, and combining this information with the
expansion velocity, once the Doppler shifts have been measured and
deprojected according to the adopted 3-D morphology. It has been
applied to several relatively close PNe (NGC~6543, Reed et
al. \cite{Re99}; IC~2448, NGC~6578, NGC~6884, NGC~3242 and NGC~6891,
Palen et al. \cite{Pa00}) and to one symbiotic nebula (Hen~2-147,
Santander-Garc\'\i a et al. \cite{Sa07a}).

Named the Southern Crab after its resemblance to the legs and claws of
that animal, the nebula around Hen~2--104 (also, PN~G315.4+09.4, and
V852~Cen) was discovered by Schwarz et al. (\cite{Sc89}). Although
similar, in morphology and kinematics, to several bipolar planetary
nebulae (Corradi \& Schwarz \cite{Co93b}), the system has been
unequivocally recognized as a genuine symbiotic star after Whitelock 
 (\cite{Wh87}) detected a Mira in its core by
means of long-term photometric monitoring.  The dynamics of this
magnificent nebula, which spans 70''$\times$35'' in the plane of the
sky, has been studied by Corradi \& Schwarz (\cite{Co93a}) and then
analysed in great spatial and spectral detail by Corradi et
al. (\cite{Co01}).

The distance to Hen~2--104 is very controversial. While Wright and
Allen (\cite{Wr78}) found a distance of 7.92~kpc, Schwarz et
al. (\cite{Sc89}) estimated 0.8~kpc.  Whitelock (\cite{Wh87}) and then
Corradi et al. ({\cite{Co01}) used the Period--Luminosity relation
for Mira stars to derive an intermediate distance of 4.7~kpc and
4.4~kpc, respectively.

In this paper, we present the results of the analysis of an NTT+SOFI
NIR spectrum, integral-field high-resolution VLT spectra, as well as
of an 2003 HST [N{\sc ii}] image of the nebula which is compared to
the one obtained on 1999.  With these, we present the first
spectroscopic detection of the Mira, the determination of the
expansion parallax of the nebula, its density map, and an estimation
of its ionized mass and total luminosity of the system.

\section{Observations and Data reduction}

\subsection{Imaging}

\begin{figure*}
\sidecaption
\resizebox{9cm}{!}{\includegraphics{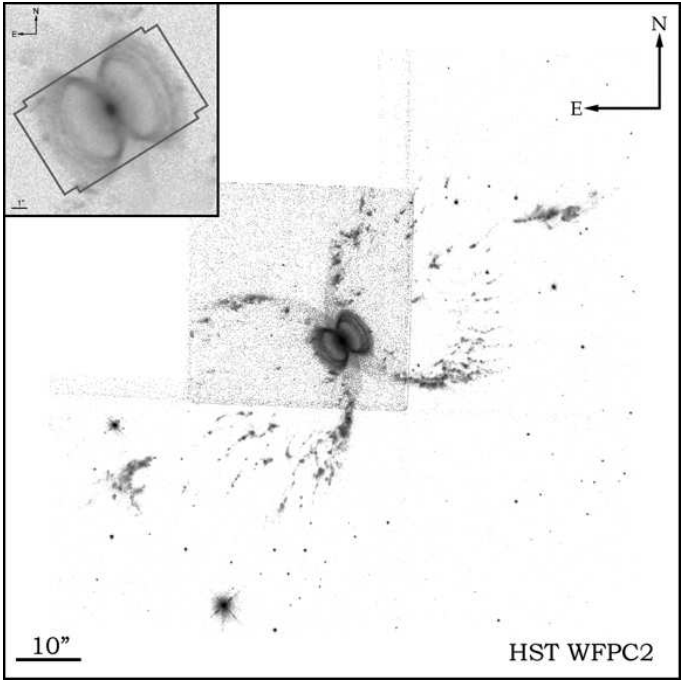}}
\caption{The HST F658N image of the nebula of Hen~2--104 obtained in 2003.
The ARGUS IFU unit covering the inner
lobes is overlaid on the top--left zoomed region.}
\label{F1}
\end{figure*}

Two sets of [N{\sc ii}] 658.3 nm images of Hen~2--104 were obtained
using the Wide Field Planetary Camera 2 (WFPC2) on the Hubble Space
Telescope (HST) on May 22, 1999 (programme GO7378) and May 10, 2003
(programme GO9336) respectively (see Fig.~\ref{F1}). Images from both
epochs were obtained through the F658N filter (central
wavelength/bandpass = 659.0/2.2 nm), which, assuming no significant
degradation between epochs, transmits the [N{\sc ii}] line with 78\%
efficiency and the H$\alpha$ line with 1\% efficiency at the
heliocentric systemic velocity of Hen~2--104. An additional set of WFPC2 
images was taken in May 20, 2001 (programme GO9050), in the F656N filter (central
wavelength/bandpass = 656.4/2.2 nm). In this case, the transmission efficiency is 
78\% in H$\alpha$, 11\% in [N{\sc ii}] 654.8 and 6\% in [N{\sc ii}] 658.3.

 We computed the observed total fluxes from the nebula in both H$\alpha$ and 
[N{\sc ii}] 658.3~nm  following 
the procedure described by Luridiana et al. (\cite{Lu03}). After subtraction of the central unresolved core and decontamination from continuum and neighbour emission lines, the 
observed total fluxes from the nebula were estimated to be 
3.6 10$^{-12}$ in H$\alpha$ and 2.1 10$^{-12}$ erg~cm$^{-2}$~s$^{-1}$ in 
[N{\sc ii}] 658.3~nm. These are consistent with the values given by Shaw \& Kaler (\cite{Sh89}) 
for a 10$''$ diameter aperture centered on the star.



The HST images (one dataset per epoch) were obtained using a 4-point
dithering pattern in order to properly sample the HST point spread
function (PSF). They were then reduced using the ``multidrizzle
package'' (Koekemoer et al. \cite{Ko02}), an automated dithered image
combination and cleaning task in PyRAF (Python IRAF).  This routine
first creates a bad-pixel mask using calibration reference files, and
performs the sky subtraction. Next, the software arranges the four
WFPC2 CCDs into a mosaic, where geometric distortion effects are
corrected, including time dependent distortions due to CCD rotations
and flexure. Then it removes cosmic rays and creates, for each epoch,
a combined and drizzled image with a sampling scale of 0$''$.045
pix$^{-1}$.





In order to extract the expansion on the sky of Hen~2--104 from 
  these two images, they need to be brought to exactly the same 
  astrometric grid. This is not the case due to slightly different 
  roll angles and pointings, and possible residuals in the distortion 
  correction done with multidrizzle for the two epochs.
  We used THELI\footnote{ftp.ing.iac.es/mischa/THELI/} 
  (Erben et al. \cite{esd05}) to define an astrometric reference frame based on the
  stellar sources in the field. Given the small rotation and distortion 
  corrections, the images were then resampled to the new grid, using the
  coordinates of the central star of Hen~2--104 as the reference projection
  point. The relative astrometric accuracy between the two images is
  on the order of 0.2 pixel.


\subsection{Integral Field Spectroscopy}


Integral Field Spectroscopy was obtained at Kueyen, Telescope Unit 2
of the 8~m VLT at ESO's observatory on Paranal, on April 2, 2004. The
inner lobes of the nebula were covered by the ARGUS IFU at
PA=122$^\circ$ using the lowest spatial resolution (314 fibres each
with a size of $0''.52$ projected onto the sky, giving a total of 
11$''$.4 by 7$''$.3 field of view).  Two 300~s exposures were taken
with the H679.7 grating, covering the region between 660 and 690 nm,
and then co-added. The resolving power was R=30000, corresponding to a
line FWHM of 10.5 km~s$^{-1}$.  The seeing was 0$''$.7.  The VLT data were
kindly reduced by Reinhard Hanuschik through the ESO pipeline, and
then further analysed using standard IRAF routines.

\subsection{Near-infrared spectroscopy}

\begin{figure*}
\center
\resizebox{18cm}{!}{\includegraphics{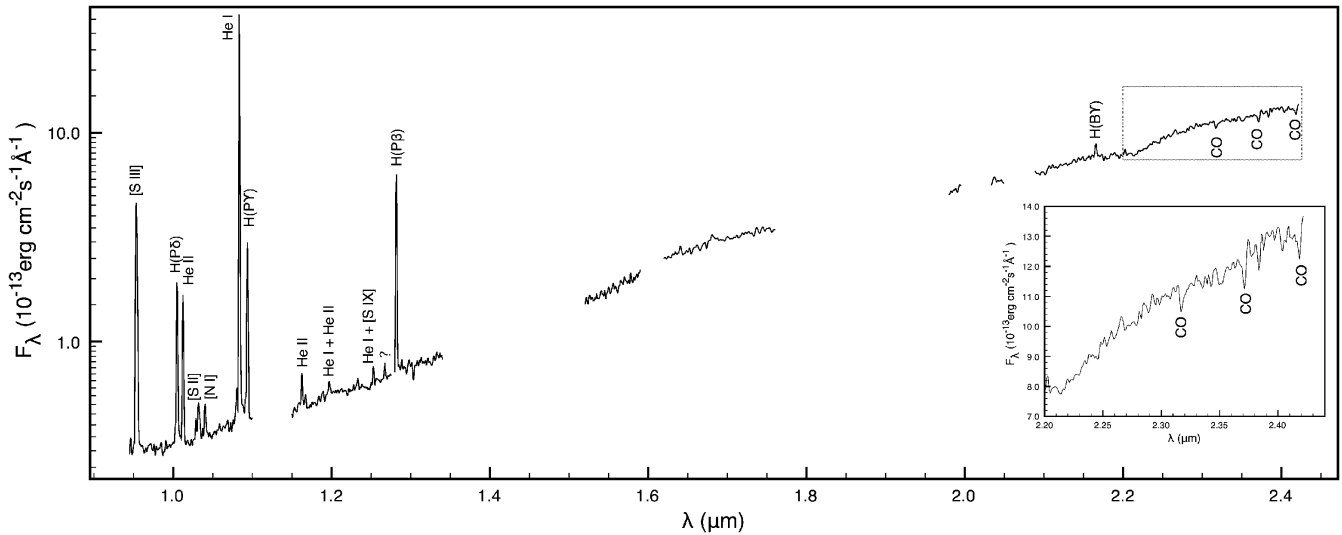}}
\caption{NTT+SOFI NIR spectrum of the core of Hen~2--104. Regions were
telluric absorption bands severely degraded the spectrum have been
removed. {\bf Right bottom:} Magnification of the region spanning from 2.2 
to 2.45 $\mu$m.}
\label{FNIR}
\end{figure*}

A 120~s near-IR spectrum of the core of Hen~2--104 was taken with SOFI
at the ESO New Technology Telescope (NTT) on May 2, 2004, in the
0.95-2.5 $\mu$m\ range, with a resolving power of $R$=930 for a slit
width of 0$''$.6.  The spectrum was reduced following standard IRAF
procedures.




\section{An improved spatio-kinematical modelling}

The image of the nebula surrounding Hen~2--104 is shown in
Fig.~\ref{F1}. It consists of two pair of nested, emitting gas
lobes. According to Corradi et al. (\cite{Co01}), the two pairs of
lobes are likely coeval, and not the result of two similar consecutive outbursts (with
the inner lobes being a younger version of the outer ones) since their
expansion velocities are (dramatically) different, and their
kinematical ages are very similar. Also the polar jets of the nebula are
roughly coeval with the inner and outer lobes.  The inner lobes show
latitude dependent surface brightness variations, which produce the
systems of ellipses observed in the HST images (Corradi et
al. \cite{Co01}).
The outer lobes, on the other hand, appear fragmented by
 Kelvin-Helmholtz instabilities into a myriad of clumpy and filamentary
structures. Most of them, specially in the furthest regions, have
outward facing tails (see Contini \& Formiggini \cite{CoFo01};  Gonz\'alez, Raga \& Steffen \cite{Go05}).

 We used the VLT IFU to map the 2-D distribution and Doppler shifts
of the emission from the core
and the whole inner lobes. The core shows a bright He{\sc i}\
$\lambda$667.8 emission line, which is also visible throughout the inner
nebula. The [S{\sc ii}] \
$\lambda\lambda$671.7,673.1 emission lines present a high S/N ratio
over the whole inner lobes, while are somewhat weaker in the portion
of the outer lobes contained in the small field of view of the
IFU. This, together with a wavelength dispersion large enough to
separate the emission from the two system of lobes wherever their
respective radial velocities differ in more than 40 km~s$^{-1}$,
allows us to make a 3-D analysis of the two structures separately.

The IFU [S{\sc ii}] 22$\times$14 fiber-array was rearranged in columns
and rows to construct a grid of pseudo long-slit spectra, with length
11.$''$4 and 7.$''$28 respectively, and slit width 0.$''$52. A
spatio-kinematical model was then built and fit to the spectral data
and to the shape of the inner lobes and base of the outer lobes in the
[N{\sc ii}] HST image (see Santander-Garc\'\i a et al. \cite{Sa04} for
a detailed description of the method).  An analytical description
assuming axial symmetry and using the formula in Solf \& Ulrich (\cite{So85}) was
adopted, with a surface, described in spherical coordinates by:

$$
r = tD^{-1} [v_{equator} + (v_{polar} - v_{equator}) \sin{|\theta|}^\gamma]
$$

where $r$ is the apparent distance to the central source,
$tD^{-1}$ the kinematical age of the outflow per unit of distance to the nebula, $v_{polar}$ and
$v_{equator}$ the velocities of the model at the pole and equator
respectively, $\theta$ the latitude angle of the model, and $\gamma$
a dimensionless shaping factor. This
assumes that each gas particle travels in the radial
direction from the central star with a velocity proportional to its
distance to the central source.  For the inner lobes, emission is also
restricted to the latitudes that correspond to the observed system of
ellipses.  In both cases, the resulting two-dimensional model was
scaled to fit the size of the object, rotated around the symmetry axis
to produce a three dimensional figure, and inclined to the plane of
the sky. The resulting geometrical shape and velocity field is used to
generate simplified model images and velocity plots for direct
comparison with the real image and spectra.

\begin{figure}
\resizebox{4cm}{!}{\includegraphics{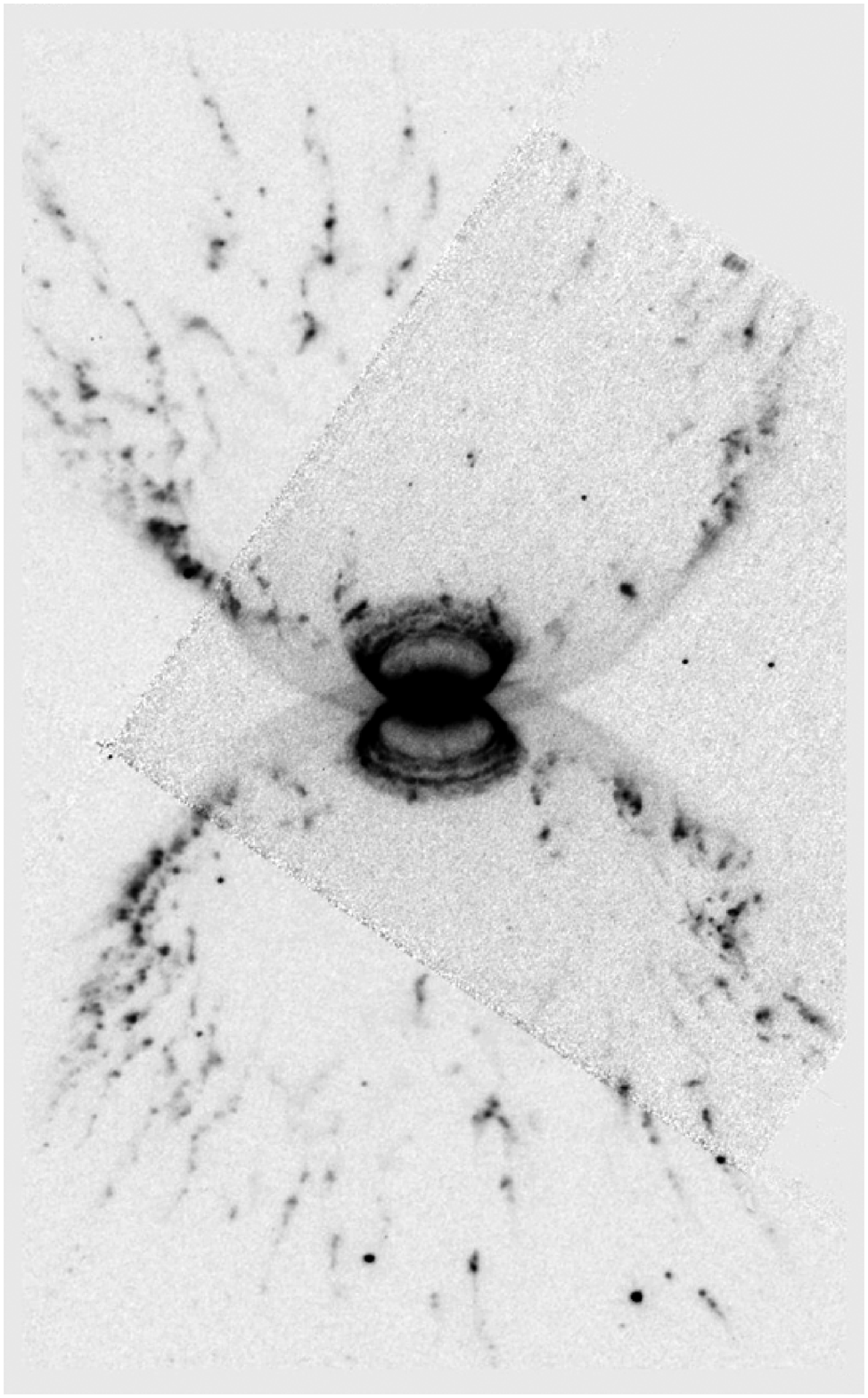}}
\resizebox{4cm}{!}{\includegraphics{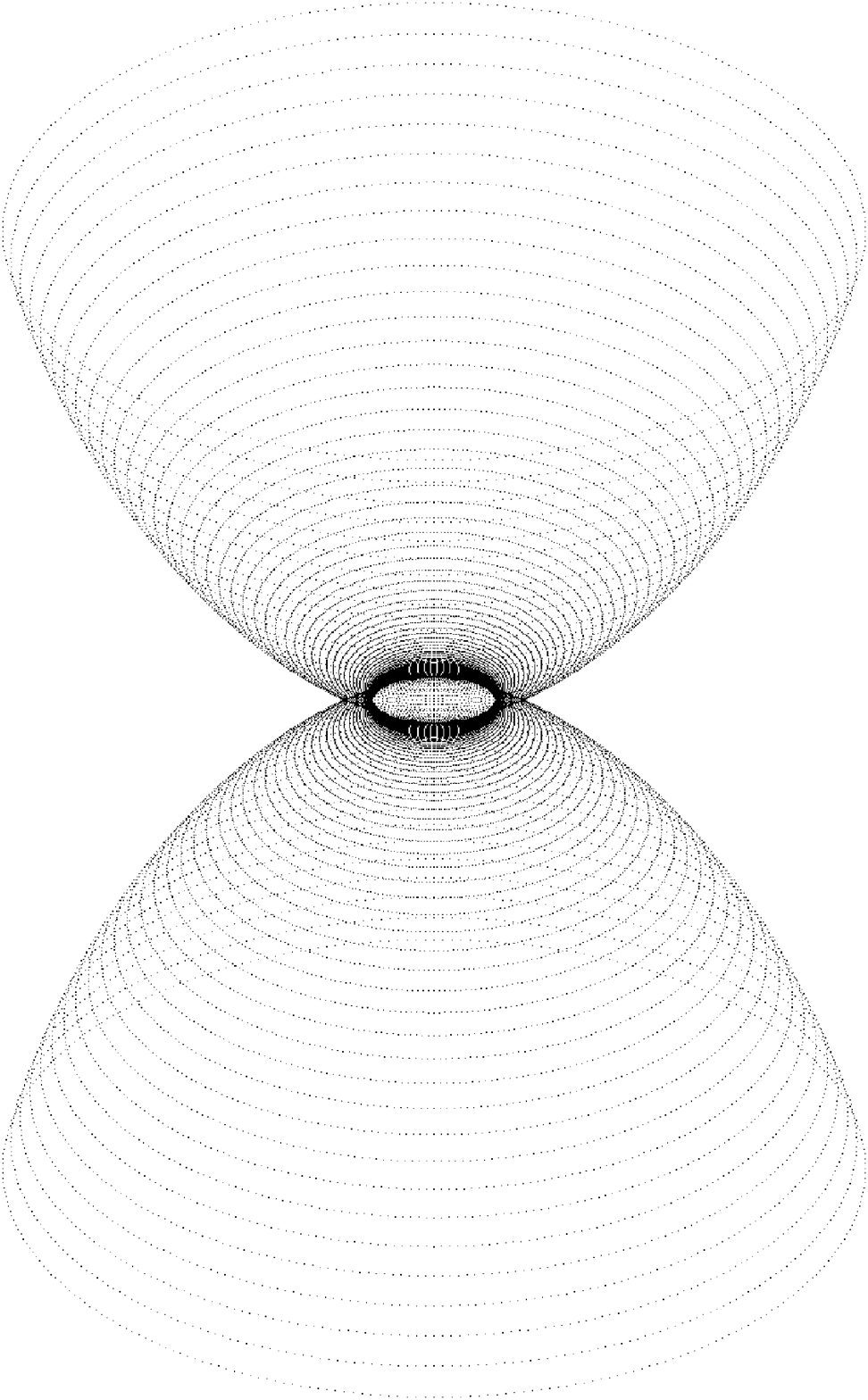}}
\caption{{\bf Left:} The [N{\sc ii}] HST image of the inner and outer
lobes of Hen~2--104. {\bf Right:} Adopted model of the outer lobes.}
\label{Fouterskima}
\end{figure}

\begin{figure*}
\center
\resizebox{16cm}{!}{\includegraphics{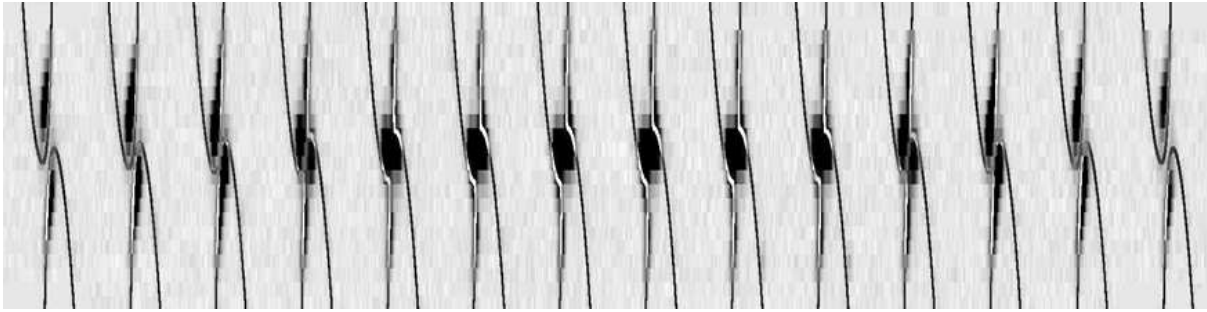}}
\caption{The ESO [S{\sc ii}] ARGUS IFU spectra rearranged so as to
simulate long-slit spectra at P.A.=122$^\circ$ (long axis of the
nebula), and offset from -3$''$.38 (SW) to +3$''$.38 (NE) from left to
right, in increments of 0$''$.52. The spatio-kinematical model of the
outer lobes has been superimposed.  Each frame is 10$''$.92 tall and
260 km~s$^{-1}$ wide. Southeastern side is up. }
\label{Fouterskspc}
\end{figure*}

\begin{table}[!h]
\begin{center}
\begin{tabular}{c c c}
Parameter & Value & Range\\
\noalign{\smallskip}
\hline\hline
\noalign{\smallskip}
$tD^{-1}$ (yrs~kpc$^{-1}$) & 1300 & (1200-1400)\\
$v_{polar}$ (km~s$^{-1}$) & 230 & (210-250)\\
$v_{equator}$ (km~s$^{-1}$) & 12 & (8-15)\\
$\gamma$ & 3.2 & (2.8-3.8)\\
$\theta_{max}$ ($^\circ$) & 52 & $(50-53)$\\
i ($^\circ$) & 58 & $(56-60)$\\
\noalign{\smallskip}
\hline
\end{tabular}
\end{center}
\label{Tsk1}
\caption{Best-fitting parameters for the outer lobes of Hen~2--104.}
\end{table}

\begin{figure}
\resizebox{4cm}{!}{\includegraphics{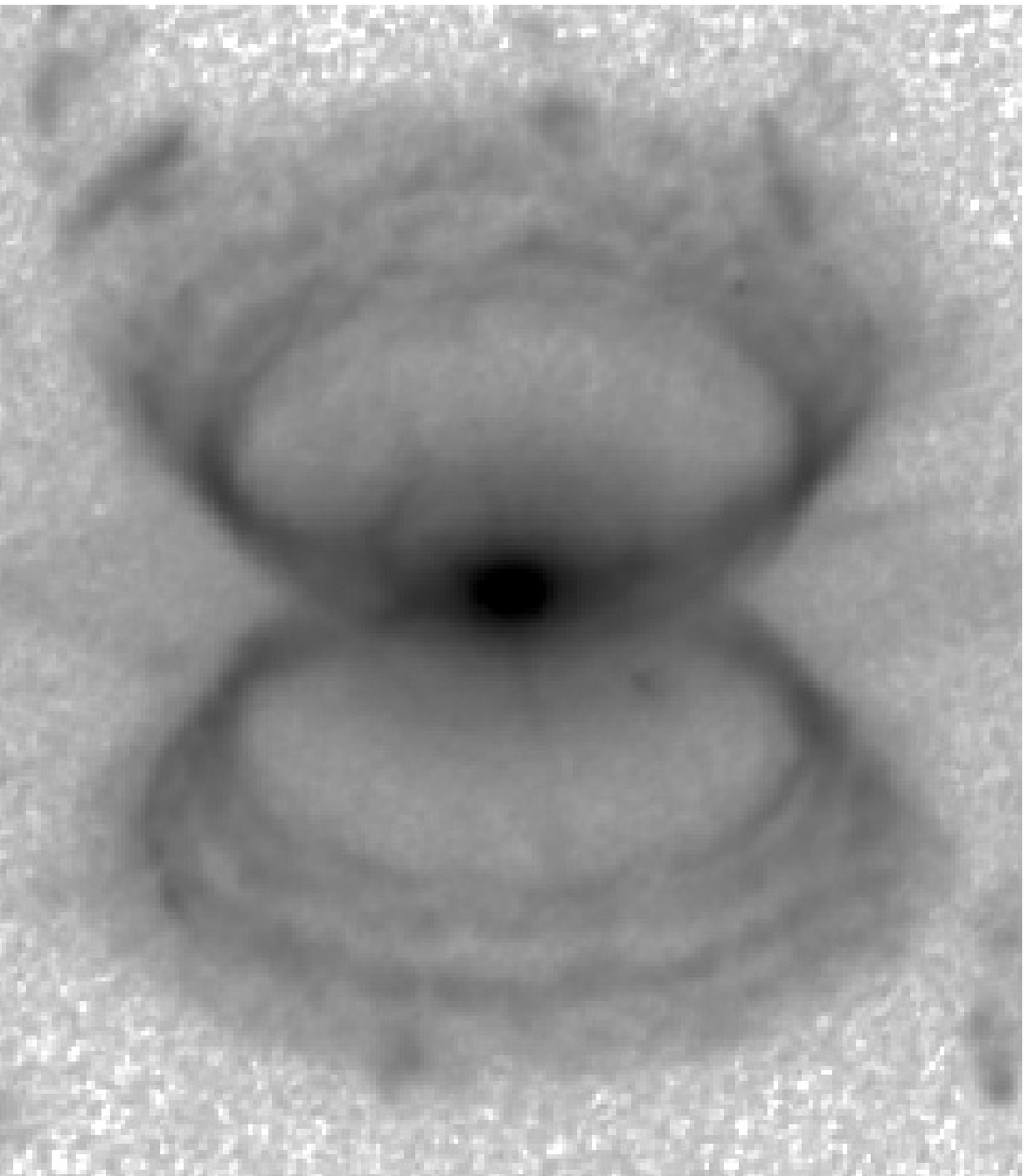}}
\resizebox{4cm}{!}{\includegraphics{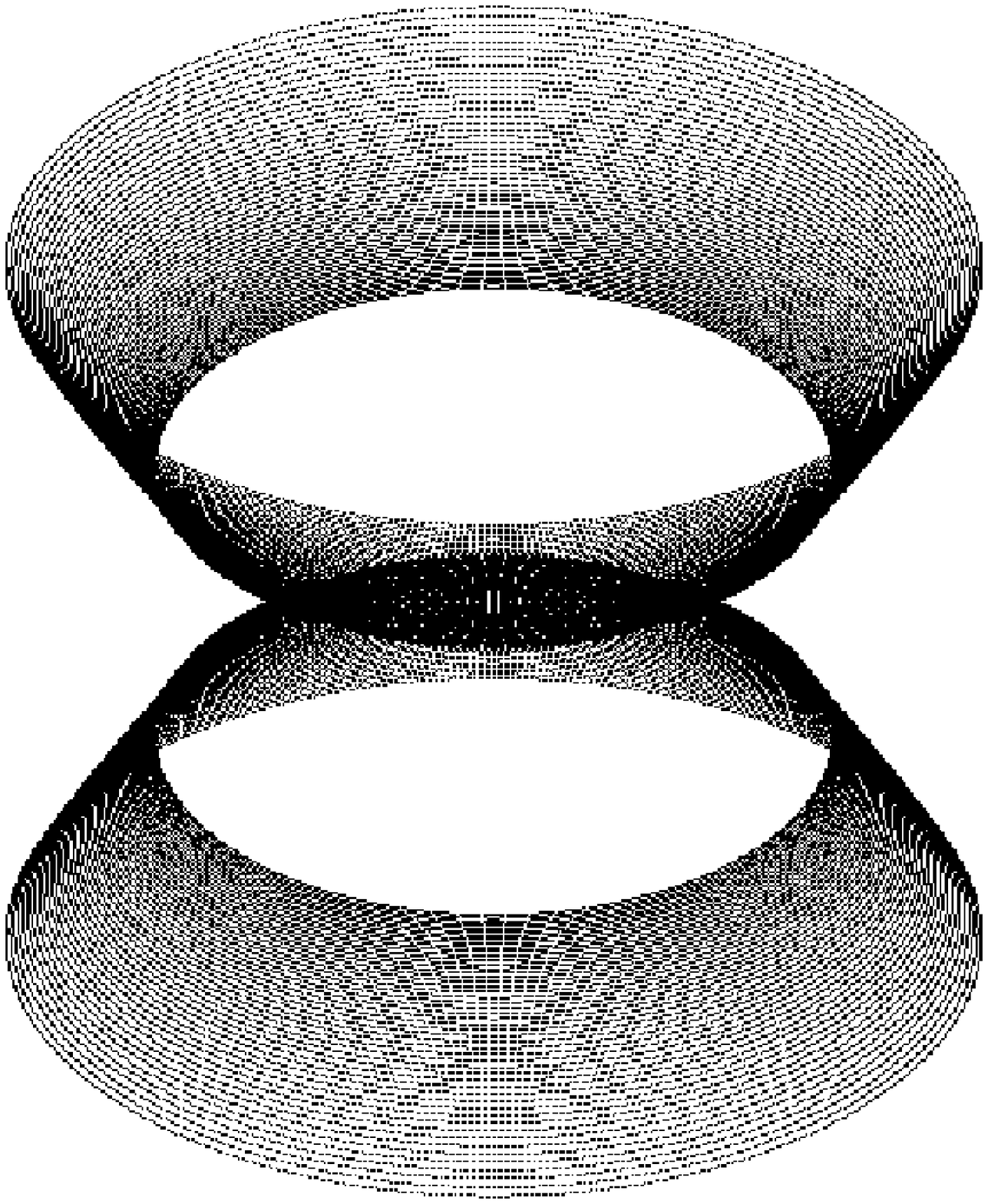}}
\caption{{\bf Left:} The [N{\sc ii}] HST image of the inner lobes of
Hen~2--104. {\bf Right:} Adopted model of the inner lobes.}
\label{Finnerskima}
\end{figure}

\begin{figure*}
\center
\resizebox{16cm}{!}{\includegraphics{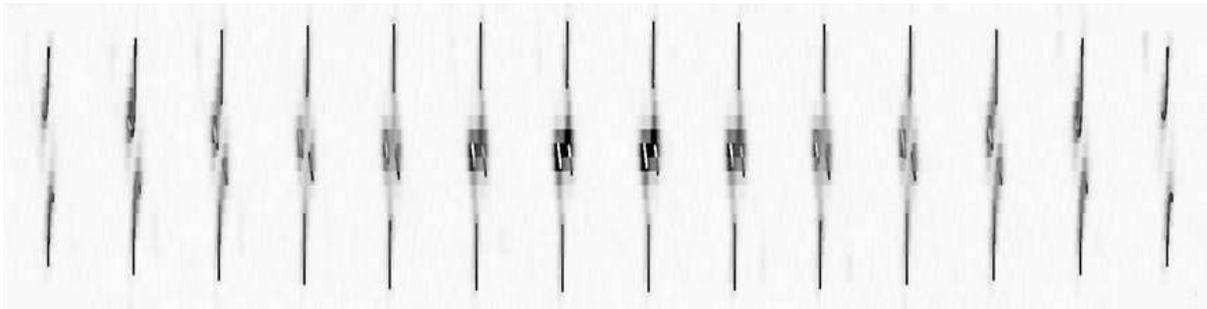}}
\caption{The ESO [S{\sc ii}] ARGUS IFU spectra arranged in columns
along the IFU longer axis. Frames correspond, from left to right, to
P.A.=122$^\circ$ and offsets from -3''.38 (SW) to +3''.38 (NE) from
left to right, in increments of 0''.52. The spatio-kinematical model
of the inner lobes has been superimposed. {\bf Note:} Each frame is 10''.92 tall and 260 km~s$^{-1}$
wide. Southeastern side is up.}
\label{Finnerskspc}
\end{figure*}

\begin{table}[!h]
\begin{center}
\begin{tabular}{c c c}
Parameter & Value & Range\\
\noalign{\smallskip}
\hline\hline
\noalign{\smallskip}
$tD^{-1}$ (yrs~kpc$^{-1}$) & 1100 & (900-1300)\\
$v_{polar}$ (km~s$^{-1}$) & 90 & (210-250)\\
$v_{equator}$ (km~s$^{-1}$) & 11.3 & (7-15)\\
$\gamma$ & 4.5 & (3.5-5.0)\\
$\theta_{min}$ ($^\circ$) & 25 & $(24-27)$\\
$\theta_{max}$ ($^\circ$) & 39 & $(37-40)$\\
i ($^\circ$) & 58 & $(56-62)$\\
\noalign{\smallskip}
\hline
\end{tabular}
\end{center}
\label{Tsk2}
\caption{Best-fitting parameters for the inner lobes of Hen~2--104.}
\end{table}

The fit to the data is performed by allowing the different parameters
of the model to vary over a large range of values and visually
comparing each resulting model to the image and spectra, until the
optimal model is found. The results are shown in
Figs.~\ref{Fouterskima}, \ref{Fouterskspc}, \ref{Finnerskima} and
\ref{Finnerskspc} and the corresponding parameters in Tables
1 and 2.

Almost all the emission from the inner lobes, both in the [N{\sc ii}]
image and the [S{\sc ii}] spectra, was found to be faithfully accounted
for by a series of rings inscribed on the walls of a truncated Solf \&
Ulrich (\cite{So85}) model at the highest latitudes, from 25$^\circ$
to 39$^\circ$, and expanding at a maximum velocity of $\sim$25
km~s$^{-1}$. The outer lobes, on the other hand, were found to share
the symmetry axis of their smaller counterpart, 58$\pm$2$^\circ$
inclined with respect to the line of sight, and with a similar, within
uncertainties, age of the nebula. We thus confirm the overall results
by Corradi et al. (\cite{Co01}), while refining the value of the
fitting parameters for the inner lobes thanks to the complete coverage
of this region of the nebula by the ARGUS IFU. A systemic velocity
v$_{LSR}$ = -85 km~s$^{-1}$ was derived, in agreement with the
measurement of Corradi \& Schwarz (\cite{Co93a}).

\section{The distance}

The distance to the nebula of Hen~2--104 has been determined by means
of its expansion parallax. This method calculates the distance by
computing the apparent expansion in the plane of the sky between two
images taken in two different epochs, and combining it with a 3-D
model of the expansion pattern, in turn usually recovered from the
analysis of the images and the radial velocity field.

The apparent expansion of both the outer lobes and polar jets of
Hen~2--104 is immediately visible by blinking the aligned 1999 and 2003
HST images. The amount of expansion of the outer lobes and jets was
quantified in two different ways (Reed et al. \cite{Re99}): 1)
deriving the growth of the whole outer lobes and jets with the
so-called magnification method; and 2) working out the expansion of
several individual knots via cross-correlation of the corresponding
surface brightness radial profiles.

\subsection{Expansion parallax via the magnification method}


A residual image of the nebula was obtained by 
subtracting the 1999 image from the 2003 one, once the images have
been registered and aligned with respect to the central core. A
generalized pattern can be seen in the jets and edges of the outer
lobes, with positive residuals on the outside and negative
ones on the inside, indicating the expansion of the nebula in
the period from 1999 to 2003.

In order to quantify the growth of the nebula, we determined the
magnification factor, $M$, to be applied to the 1999 image, which
minimizes the {\it rms} of the residual image obtained by subtracting
the magnified image from the 2003 image. In practice, though, the
complex structure of the nebula, the evolution of the brightness and
shape of individual knots and the small, non-radial components of
movement of each knot (Rayleigh-Taylor instabilities, diffusion into the medium, etc.)
 complicate the analysis.

\begin{figure*}
\sidecaption
\resizebox{9cm}{!}{\includegraphics{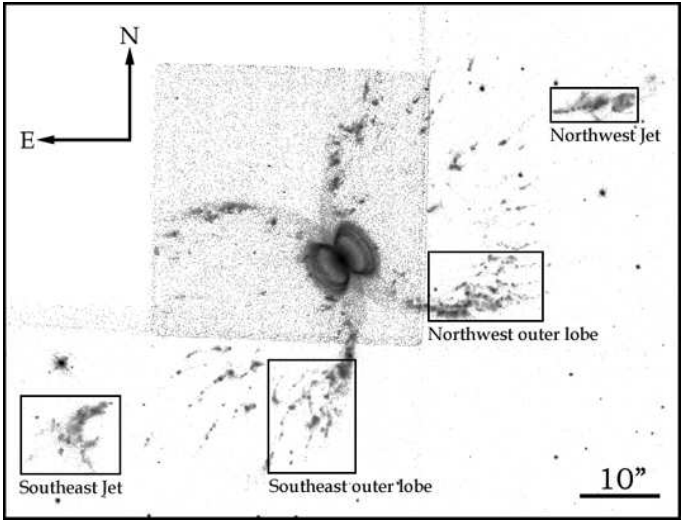}}
\caption{HST [N{\sc ii}] WFPC2 2003 image. The four regions in which
the rms of the magnification method residual image were measured are
indicated by the black boxes.}
\label{Fregiones}
\end{figure*}



Each jet, as well as two well exposed regions of the outer lobes
(see Fig.\ref{Fregiones}), were independently analysed.
The minimum {\it rms} of the residual images of the Northwest and
Southeast jets were found to correspond to magnification factors $M$
equal to 1.0009 and 1.0010 respectively; both with an uncertainty of
$\pm$0.0002. On the other hand, the $M$ factors which minimized the
residual images of the Northwest and Southeast regions of the outer
lobes were 1.0008 and 1.0012 respectively, while the uncertainty in
these cases was $\pm$0.0003. Independent comparisons of the residual
images by eye confirmed all these results. Thus, the growth of both
the jets and outer lobes is uniform, within the errors, with average
values of $M$=1.00095$\pm$0.00020 and $M$=1.00100$\pm$0.00030,
respectively.

Given the magnification factor $M$ for the jets and outer lobes, and
the kinematical age-distance parameter tD$^{-1}$ (1300 yrs~kpc$^{-1}$
for the outer lobes, from Tab.~\ref{Tsk1}, and 1200 yrs~kpc$^{-1}$ for the
jets, Corradi et al. \cite{Co93a}), the expansion parallax can be
determined by comparing the amount of growth of each feature in the
lapse of time between 1999 and 2003, $t_B$, with the features'
kinematical age. Hence:

$$
D(kpc) = \frac{t_B (yrs)}{(M-1)\ tD^{-1} (yrs\ kpc^{-1})}.
$$

We then found that the magnification method applied to the jets and outer lobes
of Hen~2--104 gives consistent values of $D_{jets}$=3.5$\pm$0.7~kpc
and $D_{outer lobes}$=3.1$\pm$0.9~kpc, respectively.

\subsection{Expansion parallax via radial profile method}

While not providing a global, intuitive vision of the growth of the
nebula and of deviations from pure radial expansion, the method that
involves the study of radial profiles allows us to trace the radial
expansion of individual features.

Essentially, this procedure consists of the analysis of
one-dimensional surface brightness radial profiles passing through a
given bright and sharp knot. Ideally, any profile from the 2003 image
will be similar to the one from 1999, except that the feature will be
shifted outwards. In order to measure the
angular expansion $\dot{\theta\ }$of the feature in the plane of the
sky, the 1999 and 2003 profiles are cross--correlated.

\begin{figure*}
\center
\resizebox{18cm}{!}{\includegraphics{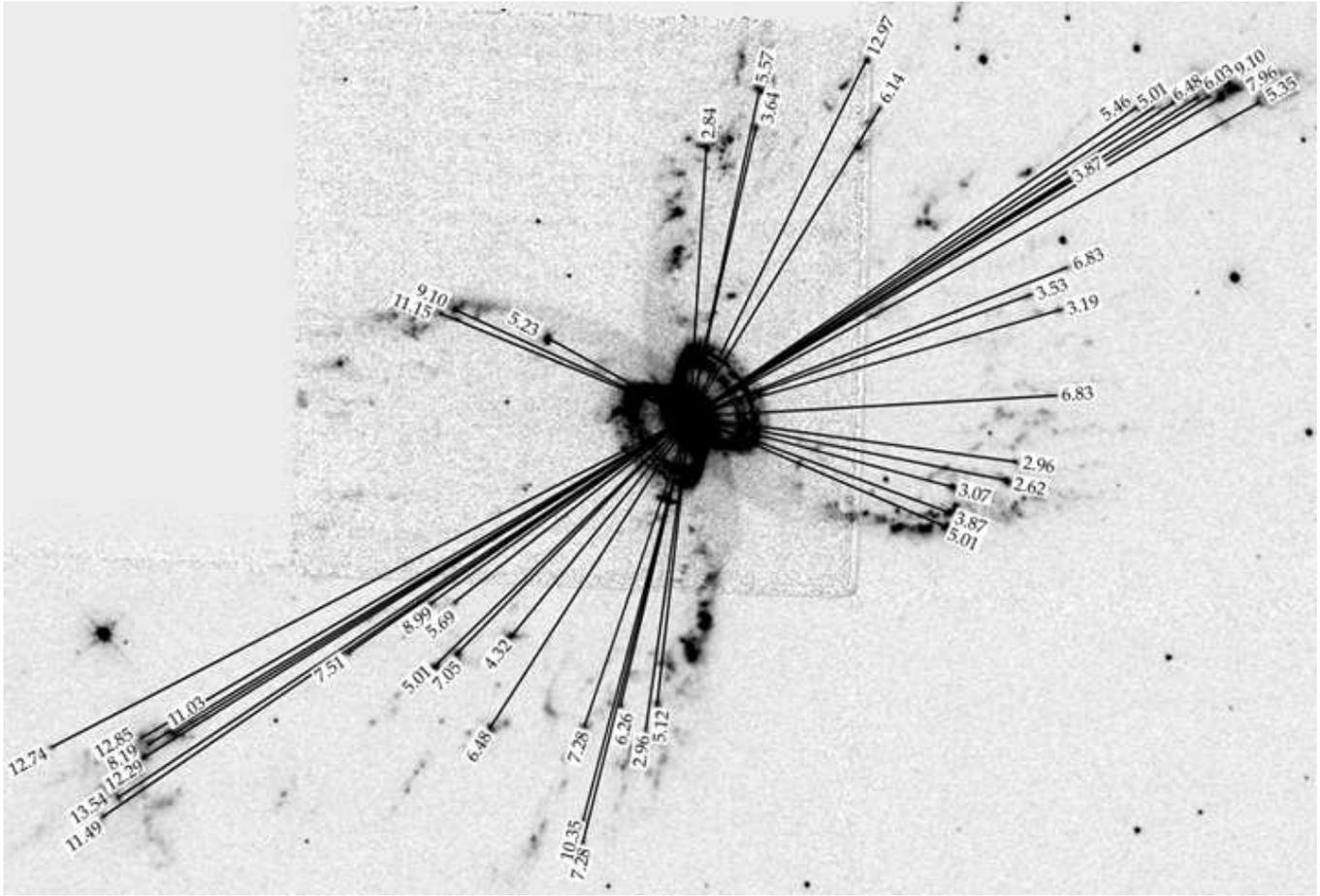}}
\caption{HST [N{\sc ii}] WFPC2 2003 image along with the radial
expansion (in mas~yr$^{-1}$) of the knots of the outer lobes and jets
analyzed in section 3.}
\label{Fradioperfiles}
\end{figure*}

Fig.~\ref{Fradioperfiles} shows the angular expansion $\dot{\theta\ }$
(in milliarcseconds) of 45 well-defined knots chosen in the outer
lobes and jets.



Assuming pure radial motion of the features, the distance to the
nebula is computed by comparing the angular expansion of each knot to
its velocity in the plane of the sky:

$$
D(kpc) = 0.211 \frac{v_{sky}(km~s^{-1})}{\dot{\theta\ }_{mas\ yr^{-1}}}
$$

For each knot in the jets, its radial velocity from the major axis
spectrum at P.A.=122$^\circ$ from Corradi et al. (\cite{Co01}), was
coupled with the expansion in the plane of the sky derived from the
cross-correlation of the radial profile. The inclination is assumed to
be the same as for the inner and outer lobes (Sect.~3).  For the outer
lobes, the velocity in the plane of the sky is computed according to
the spatio-kinematical model in Sect.~3.

After averaging the results for all the knots in the outer lobes, we
obtained $D_{outer lobes}$ = 3.1$\pm$1.0~kpc, whereas
$D_{jets}$ = 3.4$\pm$1.0~kpc for the knots in the jets.

The distance found through radial fitting is therefore fully consistent with the one
found through the magnification method; in the following, we will
adopt an average distance of $D_{Hen~2-104}$ =
3.3$\pm$0.9~kpc. Note the considerable size of the nebula at this
distance, amounting to $\sim$2.0 pc, and the large kinematical ages
of the inner and outer lobes, which would be 3700$\pm$1200, and 4200
$\pm$1200 years, respectively.

\section{The NIR spectrum}

The NIR spectrum of the core (see Fig. \ref{FNIR}) shows a variety of
emission lines over an almost featureless rising continuum.

The region between 0.95 and 1.3 $\mu$m shows Hydrogen and Helium
atomic recombination lines, as well as [SII], [SIII], [NI], and H$_2$.
The region onward 2.1 $\mu$m shows three CO absorption bands typical
of late AGB stars (e.g.~Kamath \& Ashok \cite{Ka03}). These
bands, although barely visible in the dominant continuum presumably
produced by hot circumstellar dust, show for the first time some
photospheric spectral features of the Mira in the system, so far only
detected thanks to its characteristic NIR photometric light curve
(Whitelock \cite{Wh87}).

The emission lines and absorption band CO identifications are shown in
Table 3.

 The featureless NIR continuum of Hen~2--104 is quite peculiar
for symbiotic stars (cf. Whitelock \cite{Wh03}), where in general this
part of the spectrum is dominated by deep absorption bands of the red
giant.
Note that Mz~3 and M~2-9, two nebulae suspected of having a
symbiotic binary nucleus, show a similar spectrum. It is likely that
the rising continuum is emission produced by a significant amount of
hot circumstellar dust, but a detailed modelling of the spectrum is
beyond the scope of these paper. What is instead relevant for the
present discussion, is the fact that it is clear that the emission in
the K-band is not always a good measure of the luminosity of the red
giant of the system, so that distances computed from the Mira
period-luminosity relationship using the K magnitude, are necessarily
subject to large uncertainties.



\begin{table}[!h]
\begin{center}
\begin{tabular}{c c c c}
$\lambda_0 \  (laboratory)$ & Element  & Flux  & error \\
(\AA)  &  (emission) & (10$^{-13}$  $\frac{erg}{cm^2 s})$ & (\%) \\
\noalign{\smallskip}
\hline\hline
\noalign{\smallskip}
9463.61  & He {\sc i}  3$^3$S-5$^3$P$^0$    &  2.11   & 14 \\
9532.55  & [S{\sc iii}]        & 106.0 &   1 \\
10027.73  & He {\sc i}   3$^3$D-7$^3$F$^0$   &  0.7      &  17 \\
10049.37& H(P$\delta$) &  28.6   &   1 \\
10123.61  & He {\sc ii}$_{\ \ 4-5}$           &  23.1    &  1 \\
10286.5  & [S{\sc ii}]         & 1.32     & 10 \\
10320.3  & [S{\sc ii}]          & 4.83   & 7  \\
10370.5  & [S{\sc ii}]          &  0.44   &  20 \\
10397.7+10398.2 & [N{\sc i}] & 3.62   & 7 \\
 10830.2 & He {\sc i}      & 640.0   &     1  \\
 10912.0 & He {\sc i}      & 1.85      &    18 \\
 10938.1 & H(P$\gamma$) & 44.9   &  1 \\
 11626.2 & He {\sc ii}$_{\ \ 5-7}$  &  3.85  & 10 \\
 11673.24 & He {\sc ii}$_{\ \ 6-11}$ & 1.17 & 31 \\
 11970.0 & He {\sc i}+He {\sc ii} & 2.19 & 36 \\
 12329.0 & H$_2$(3,1)S(1) & 1.8 & 18 \\
12528.0 & He {\sc i}  & 2.67 & 10 \\
12673.   &   ?     & 2.3  & 10 \\
12785.0 & He {\sc i} & 5.2 & 7 \\
12818.08 & H(P$\beta$) & 97.0 & 1 \\
16407.19 & H(Br12) & 6.98 & 39 \\
16806.52 & H(Br11) & 10.2 & 52 \\
21655.29 & H(Br$\gamma$) & 28.9 & 15 \\
\noalign{\bigskip}
\hline
\noalign{\smallskip}
$\lambda \ (band head)$ & Molecule  & Flux  & error \\
($\mu$m)  &  (absorption) & (10$^{-13}$  $\frac{erg}{cm^2 s})$ & (\%) \\
\noalign{\smallskip}
\hline\hline
\noalign{\smallskip}
2.32 & CO  & 0.34 & 28 \\
2.37 & CO  & 0.35 & 46 \\
2.42 & CO  & 0.59 & 27  \\
\noalign{\smallskip}
\hline
\end{tabular}
\end{center}
\label{T1}
\caption{Hen~2-104: NIR emission line and absorption band
identification, along with their fluxes and associated errors, which
include both the poissonian error and the error in the determination
of the continuum. ``?''  denotes that the line was unidentified.}
\end{table}

\section{Nebular density and mass}

\subsection{Density map}


\begin{figure}
\resizebox{8cm}{!}{\includegraphics{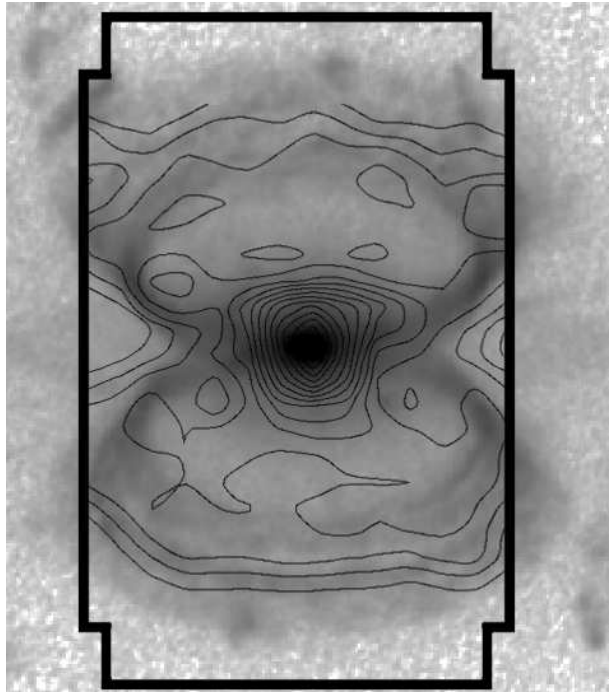}}
\caption{Density map of the inner lobes and base of the outer lobes of
Hen~2--104. The contour levels span from n$_e$ = 200 to 3000 cm$^{-3}$
with step 200.}
\label{Fdensity}
\end{figure}

The integral field spectroscopy data, featuring the [S{\sc ii}] \
$\lambda\lambda$671.7,673.1 emission line doublet, allow to
build a density map covering the whole inner lobes and the base of the
outer lobes (Fig.\ref{Fdensity}).  As mentioned previously, the
large resolving power (R=30000) of the spectra was high enough to
separate the inner lobes from the outer lobes wherever their
respective radial velocities differed more than 40 km~s$^{-1}$.

In the absence of actual temperature measurements for the lobes,
we assume $T_e$=10000 K for both structures. The resulting densities
range from $n_e$=500 to 1000 cm$^{-3}$ for the inner lobes, and from
$\sim$300 to 500 cm$^{-3}$ for the base of the outer lobes.  The core
shows the highest [S{\sc ii}] densities, $n_e \sim$2400
cm$^{-3}$.  However, the [S{\sc ii}] lines seen towards the core 
probably arise in the outer layers of a very dense circumstellar envelope, 
as e.g. in the case of the other symbiotic Mira HM~Sge (Corradi \& Schwarz
\cite{Co99}). Indeed, a core density larger than 2.5 10$^6$ cm$^{-3}$ is
estimated for Hen~2--104 from the [O{\sc iii}] lines
(Freitas Pachecho \& Costa \cite{Fr96})}. At these densities, the
[S{\sc ii}] lines are collisionally quenched. These results confirm
those given by Corradi \& Schwarz (\cite{Co93a}).



\subsection{Ionized mass}

The ionized mass of the nebula around Hen~2--104 can be estimated via the formula

$$ M_{H^+} = \frac{m_H F(H\beta) 4 \pi D^2}{h \nu_{\beta}
\alpha^{eff}_{B} n_e}
$$

where $D$ is the distance, $n_e$ the nebular electron density,
F(H$\beta$) is the total H$\beta$ flux, $m_H$ is the mass of the
hydrogen atom, $h \nu_\beta$ the energy of an H$\beta$ photon, and
 $\alpha^{eff}_B$ the effective recombination coefficient for case B 
(adopted from Osterbrock \cite{Os89}). After correcting
the observed H$\alpha$ flux for an extinction of A$_v$=3.0~mag
(average of the values found by Lutz et al. \cite{Lu89} and Freitas
Pacheco \& Costa \cite{Fr96}), and applying the theoretical H$\alpha$
to H$\beta$ ratio of 2.85 (for an electron temperature of 10000~K), we
obtain  F(H$\beta$)=1.3~10$^{-11}$~erg~cm$^2$~s$^{-1}$. Assuming an
average density of 500~cm$^{-3}$ and the distance found in this work,
we compute an ionized mass of $\sim$0.16~M$_\odot$.  The extinction 
correction is very important in the above calculation, and, if we instead 
assume an interstellar extinction of 1.6~mag, as derived from
the Galactic extinction model by Drimmel, Cabrera-Lavers \&
L\'opez-Corredoira (\cite{Dr03}) for the location of Hen~2--104 and
the computed distance,
the resulting ionized mass is 0.05 M$_\odot$. These have to be
considered as lower limits, as they neglect any local extinction due to
dust around and inside the nebula.

 A different mass estimate can be obtained by computing, both for the 
inner and outer lobes, the volume of each
feature occupied by gas in the following manner.  As the geometry of
the surface of the lobes is known from our spatio-kinematical
modelling (Sect.~3), and the electron density distribution is also
known from the [S{\sc ii}] line ratio, the only other parameter needed
to compute the volume occupied by the emitting gas is the thickness of
the walls of the lobes, if the filling factor (cf. Pottasch
\cite{Pobook}) is assumed to be around unity. 3-D models with
different thickness have been produced using IDL (RSI) routines
adapted from the VISNEB tools developed for the Cloudy~3D code
(Morisset \cite{Mo06}). A cubic grid of 300$^3$ cells was used to
generate and rotate a 3D emissivity model (with the emissivity per
unit volume $\propto {n_e}^2$). The emission along the line of sight
was then integrated in order to generate a synthetic image in
collisionally excited lines (this also assumes that the ionic
abundance and electron temperature are constant throughout the lobes).
This procedure was repeated several times allowing the
thickness of the 3D model to vary over a wide range.

\begin{figure*}
\center
\resizebox{18cm}{!}{\includegraphics{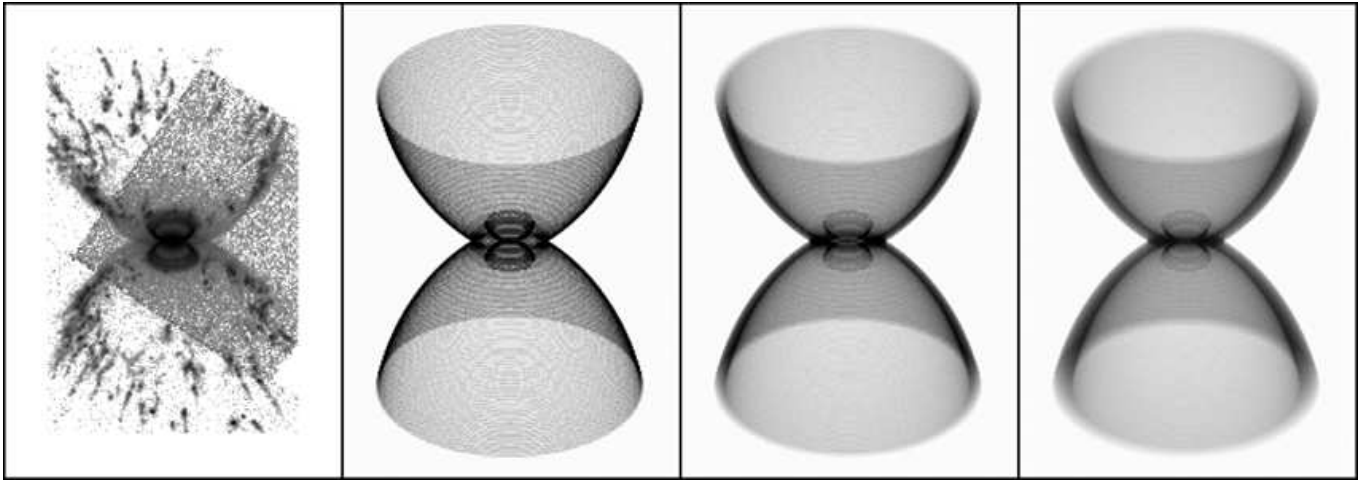}}
\caption{{\bf From Left to Right:} HST [N{\sc ii}] image from 2003 and illustrative sample of synthetic images of the nebula. The thickness of the outer lobes of the three models are 1''.0, 3''.5 and 6''.0, respectively, while the inner lobes show a constant thickness of 0''.9.}
\label{Fsinteticas1}
\end{figure*}

\begin{figure*}
\center
\resizebox{18cm}{!}{\includegraphics{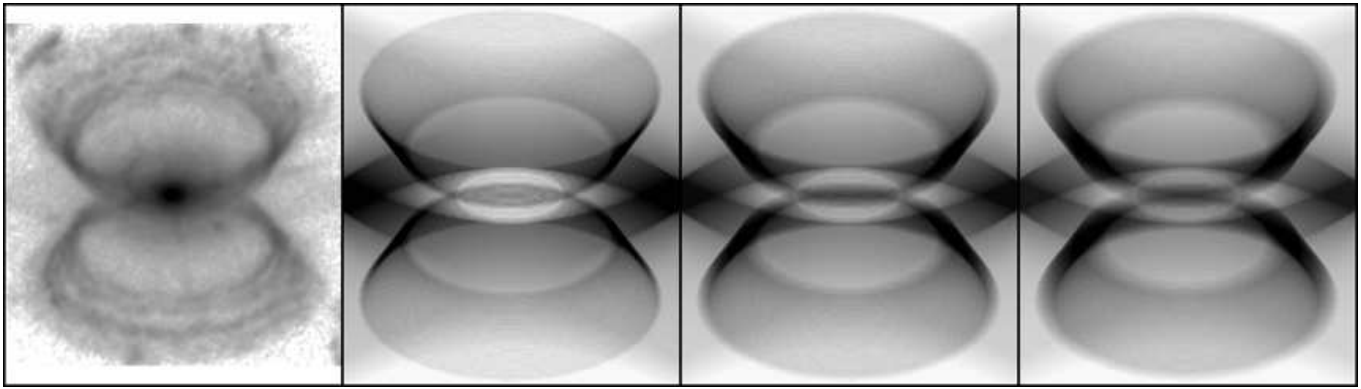}}
\caption{{\bf From Left to Right:} HST [N{\sc ii}] image from 2003 and illustrative sample of synthetic images of the inner lobes (and the central region of the outer ones). The thickness of the inner lobes of the three models are 0''.4, 0''.9 and 1''.4, respectively, while the outer lobes show a constant thickness of 3''.5.}
\label{Fsinteticas2}
\end{figure*}

In order to find the optimal thickness that accounts for the observed
inner and outer lobes, residual images were obtained by subtracting
each synthetic image from the real HST one. The regions around the
edges of each lobe (i.e. where the walls of the structure are
perpendicular to the plane of the sky and therefore its thickness is
easier to estimate) were analysed in the residual images, both by
visual comparison and by measuring the {\it rms} of each
region. Illustrative samples of these synthetic images, both for the
outer and inner lobes, are shown in Figs.\ref{Fsinteticas1} and
\ref{Fsinteticas2}.  The resulting best-fit thickness is
0$''$.9$\pm$0$''$.4 and 3$''$.4$\pm$0$''$.5 for the inner and outer
lobes respectively. Finally, in order to account for the knotty
appearance of the outer lobes, we computed the fraction of the total
volume of the model occupied by gas. The procedure to do this was to
count the number of pixels in the HST images whose surface
brightness is larger than 3 times the statistical noise of the
background, and compute which fraction of the area projected in the
sky of the model nebula is effectively filled.  The result was that
0.34$\pm$0.06 of the total volume of the outer lobes is filled by gas.

Therefore, using typical densities of n$_e$=600 cm$^{-3}$ and
n$_e$=350 cm$^{-3}$ for the inner and outer lobes, respectively, and
assuming a roughly solar abundance for the nebula, the resulting
ionized mass for the two pairs of lobes is (4.6$\pm$0.9) 10$^{-3}$ and
0.08$\pm$0.01 M$_{\odot}$, respectively,  for a total of roughly
$\sim$0.1 M$_\odot$ for the whole nebula. This value is midway, but 
broadly consistent, with the two previous estimates, 0.16 and 
0.05~M$_\odot$ obtained varying the adopted extinction, and is clearly 
pointing to a relatively large mass for the nebula.

\section{Discussion}

Schwarz et al (\cite{Sc89}) discovered the extended nebula around
Hen~2--104, and they would probably have classified it as a planetary
nebula, had  Whitelock (\cite{Wh87}) not found, two years before,
a long-period pulsation of the central star by means of NIR
photometric monitoring. In fact, apart from indirect hints suggesting
a symbiotic nature (e.g. a high density core, NIR colours unusual for
a PN, and an articulated high-excitation spectrum), this was the only
evidence for the presence of a cool giant. The NIR spectrum presented
in Sect. 5 shows for the first time the spectral signature of the red
giant in the spectrum.   However, a strong featureless continuum,
presumably from hot circumstellar dust, renders these bands
barely visible. Thus, given the difficulties of direct detection of
the Mira in some genuine symbiotic nebulae such as Hen~2--104, the
already long debate on the nature of some bipolar nebulae with dust
dominated NIR spectra and high density compact cores (e.g. Mz~3 and
M2-9, see Schmeja \& Kimeswenger \cite{Sc03}, Phillips \cite{Ph07} and
Santander-Garc\'\i a \& Corradi \cite{Sa07b})  is not a surprise.

%



 In this work, we derive a distance for Hen~2--104 of
  3.3$\pm$0.9~kpc by means of the expansion parallax method. It is
  known that the expansion parallax determinations can be affected by
  shocks. This is for example the case of the symbiotic Mira
  Hen~2--147, in which Santander-Garc\'\i a et al. (\cite{Sa07a}) found that
  distance is understimated if shocks are not properly taken into
  account. The main evidence for shocks in that nebula are the broad
  and asymmetrical line profiles, amounting to up 200~km~s$^{-1}$ full
  width at zero intensity (cf. Hartigan et al. \cite{Ha87}). This is
  not the case for Hen~2--104, where line widths in the jets and lobes
  are of the order of $\sim$15~km~s$^{-1}$, only slightly larger than
  the gas thermal broadening.  We therefore conclude that shocks do
  not strongly affect the expansion parallax determination of
  Hen~2-104, as also suggested by the consistent distance estimate for
  the lobes and the jets (shocks would a priori act differently on the
  two components)\footnote{Note, however, that even if the actual
  distance were somewhat larger, the conclusions below would still
  hold, as the ionized mass would increase accordingly.}. We consider
  our expansion parallax determination to be more robust than the
  distance obtained through the period-luminosity relationship for the
  Mira (Whitelock \cite{Wh87}), because our spectra show that the K
  band magnitude does not provide a reliable measure of the luminosity
  of the Mira.



\begin{figure}[h!]
\resizebox{9cm}{!}{\includegraphics{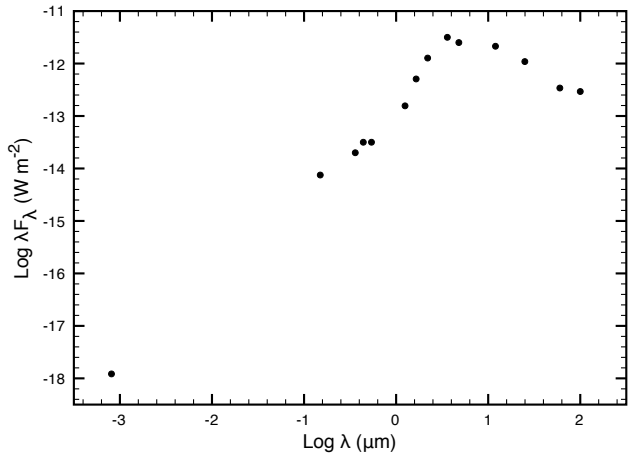}}
\caption{Observed spectral energy distribution of Hen~2--104, from
1.2~\AA\ to 100 $\mu$m, based on the values quoted in Schwarz et
al. \cite{Sc89} for the UBV and the IR photometry, the IRAS fluxes,
the IUE UV fluxes and the Chandra X ray fluxes.  }
\label{FSED}
\end{figure}

Application of period-luminosity relation (Feast et al. \cite{Fe89},
Groenewegen \& Whitelock \cite{Gr96}, and Whitelock et
al. \cite{Wh94}) to the Mira of Hen~2--104 (P=400~days, Whitelock
\cite{Wh87}), would imply a bolometric luminosity of the AGB star of
$\sim$6000 L$_{\odot}$. This can be compared with the total luminosity
of the system derived from the spectral energy distribution presented
in (Fig.~\ref{FSED}), at the computed distance of 3.3~kpc. From the
latter, in order to obtain a luminosity larger than 6000~L$_{\odot}$
(the luminosity of the white dwarf companion should also be
considered), the interstellar extinction must be $A_v>$2.9~mag. 
This value is significantly larger than the one obtained using the
Galactic model by Drimmel, Cabrera-Lavers \& L\'opez-Corredoira
(\cite{Dr03}, see Sect.~6.2), and closer to the determinations by Lutz
et al. (\cite{Lu89}), $A_v$=2.23~mag, and Freitas Pacheco \& Costa
(\cite{Fr96}), $A_v$=3.7~mag.




 Another result of this work that deserves some discussion is the
remarkable ionized mass of the extended nebula, amounting to about a
tenth of a solar mass (see section 6.2). Given such a large value, and
considering that only a negligible fraction of the Mira wind is
captured by the white dwarf, it seems clear that the main donor of
mass for the nebula is the Mira. As Corradi et al. (\cite{Co01})
suggested, this kind of nebulae are likely to be the result of the
interaction of the slow winds from the Mira with the fast winds from
an outbursting white dwarf. Given the articulated structure of these
nebulae, it is not easy to understand how the different winds
contribute to each morphological component, but it seems natural to
associate the high velocity features (like the polar jets) directly
with the fast winds from the hot white dwarf and its accretion disc.
 Many of the features of the nebula around Hen~2--104 could be
explained by the weak-jet model of the hydro-dynamic calculations by
Garc\'\i a-Arredondo \& Frank (\cite{Ga04}). Although the initial
conditions they assumed for the two case studies are different, the
conclusions of their model are applicable.  This emphasizes the key
role of the Mira in supplying the material and the essential role of
the white dwarf companion in accelerating and, in all probability,
deflecting the streamlines of the Mira's wind to shape the bipolar
morphology that we see today.
 Our spatio-kinematic fit clearly shows that the outflow speeds
are slowest at low latitudes and highest along the polar
direction. This is the same pattern as found in most bipolar PNe
(e.g. Corradi \cite{Co04}), as well as in other classes
of objects like in the homologously expanding bipolar homunculus
of $\eta$~Carina (Smith \& Gehrz \cite{Sm98}). Thus, the mechanisms that may
have formed Hen~2--104 are not limited to symbiotic stars and PNe, and
may be common when the mass loss rate of the source is high.

 We have also found that the walls of the two lobe pairs of
Hen~2--104 are very thin compared to their radial distance from the
star. A comparison to the sound crossing time allows for some
speculation about the mechanism holding the walls tightly together. If
the gas temperature is (and has been) 10$^4$~K since the ejection
$\sim$4000~yrs ago (at the computed distance), the sound speed is 10
km~s$^{-1}$, so it travels 1.2 10$^{12}$~km in 4000~yrs. The wall
thickness of the inner lobes (0$''$.9), 3.5 10$^{11}$~km, is less than
the sound travel distance, implying that something like an outside
pressure or an embedded magnetic field is holding the walls of the
inner lobes together, perhaps accounting for their relatively smooth
and structured appearance and high emission density.  The outer walls
are 1.3 10$^{12}$~km thick (3$''$.4), which is roughly the sound crossing
distance.  This means that the walls of the outer lobes are expanding
at their sound speed (i.e. driven by their own thermal pressure).


For a typical AGB wind speed of $\sim$15 km~s$^{-1}$ (Habing, Tignon
\& Tielens \cite{Ha94}), it would take $\sim$2.6 \ 10$^4$~yrs to the
gas lost by the Mira of Hen~2--104 to fill the volume presently
occupied by the nebula ($\sim$0.4 pc in radius at the adopted
distance). During this lapse of time, an average mass loss rate of
$\sim$3.5 10$^{-6}$ M$_{\odot}$ yr$^{-1}$ would be needed to
accumulate the total mass of the nebula ($\sim$0.1 M$_{\odot}$)
around the system, that later would be shaped by fast winds from the
outbursting white dwarf companion. This mass loss rate is in good
agreement with the one expected from a 6000 L$_{\odot}$ Mira variable
(Alard et al. \cite{Al01}), making the depicted scenario a coherent
one and reinforcing our distance and mass estimates.

In any case, the mass of the nebula of Hen~2--104 is remarkably larger
than previously thought for this class of objects, and more similar to
the ionized masses of genuine PNe (Pottasch \cite{Pobook}).  Thus the
Mira in Hen~2--104 would be in a quite evolved stage, close to the tip
of the AGB, where due to strong mass loss a massive circumstellar
envelope can be built up. The ionized nebula that we observe can
therefore be seen as a sort of ``premature'' PN, in which the nebular
shaping and excitation is done by the WD companion, which
``anticipates'' the post-AGB evolution of the Mira. Note that the
kinetic energy of the nebula of Hen~2--104 ($\sim$8 10$^{44}$~erg) and
its momentum ($\sim$10$^{39}$~g~cm~s$^{-1}$) are similar to those of
regular PNe, and significantly lower than the momentum computed by
Bujarrabal (\cite{Bu01}) for a number of multi-polar young and
proto-PNe. The energetics of the outflow of Hen~2--104 can be
accounted by the Mira luminosity and the energy released by a
slow-nova explosion on the WD companion.

In the past, it has been often stated that a distinctive
characteristic of the nebulae around symbiotic stars is that they are
one or two orders of magnitude less massive than PNe.  Hen~2--104
shows that this is not the rule, so that the line dividing PNe from
symbiotic nebulae might actually be more blurred than previously
thought.

\begin{acknowledgements}
The authors wish to thank Vera Kozhurina-Platais, Warren Hack, Anton Koekemoer, and Richard Hook from STScI for their kind help in multidrizzle usage, and Reinhard Hanuschik for the FLAMES dataline processing.

\end{acknowledgements}

\end{document}